\documentclass{aa}  

\usepackage{natbib}
\usepackage{graphicx}
\usepackage{txfonts}
\usepackage{xcolor}
\usepackage[colorlinks=true,citecolor=blue]{hyperref}

\makeatletter
\renewcommand*\aa@pageof{, page \thepage{} of \pageref*{LastPage}}
\makeatother

\begin{document} 
  \title{The star formation timescale of elliptical galaxies}
  \subtitle{Fitting [Mg/Fe] and total metallicity simultaneously}

  \author{Zhiqiang Yan\inst{1} \fnmsep \inst{2}
          \and
          Tereza Jerabkova \inst{1} \fnmsep \inst{2} \fnmsep \inst{3} \fnmsep \inst{4} \fnmsep \inst{5}
          \and
          Pavel Kroupa \inst{1} \fnmsep \inst{2} \fnmsep 
          }

  \institute{Helmholtz-Institut f{\"u}r Strahlen- und Kernphysik (HISKP), Universität Bonn, Nussallee 14–16, 53115 Bonn, Germany
             \\ Emails: yan@astro.uni-bonn.de; tereza.jerabkova@eso.org; pkroupa@uni-bonn.de
         \and
             Charles University in Prague, Faculty of Mathematics and Physics, Astronomical Institute, V Hole{\v s}ovi{\v c}k{\'a}ch 2, CZ-180 00 Praha 8, Czech Republic
        \and 
            Astronomical Institute, Czech Academy of Sciences, Fri\v{c}ova 298, 25165, Ond\v{r}ejov, Czech Republic
        \and 
            Instituto de Astrof{\'i}sica de Canarias, E-38205 La Laguna, Tenerife, Spain
        \and
            GRANTECAN, Cuesta de San Jose s/n, 38712 Brena Baja, La Palma, Spain
             }

  \date{Received 5 Sep 2019 / Accepted 29 Oct 2019}

  \abstract
  {
    The alpha element to iron peak element ratio, for example [Mg/Fe], is a commonly applied indicator of the galaxy star formation timescale (SFT) since the two groups of elements are mainly produced by different types of supernovae that explode over different timescales.
    However, it is insufficient to consider only [Mg/Fe] when estimating the SFT. The [Mg/Fe] yield of a stellar population depends on its metallicity. Therefore, it is possible for galaxies with different SFTs and at the same time different total metallicity to have the same [Mg/Fe]. This effect has not been properly taken into consideration in previous studies.
    In this study, we assume the galaxy-wide stellar initial mass function (gwIMF) to be canonical and invariant. We demonstrate that our computation code reproduces the SFT estimations of previous studies where only the [Mg/Fe] observational constraint is applied.
    We then demonstrate that once both metallicity and [Mg/Fe] observations are considered, a more severe "downsizing relation" is required. This means that either low-mass ellipticals have longer SFTs (> 4 Gyr for galaxies with mass below $10^{10}$ M$_\odot$) or massive ellipticals have shorter SFTs ($\approx 200$ Myr for galaxies more massive than $10^{11}$ M$_\odot$) than previously thought. This modification increases the difficulty in reconciling such SFTs with other observational constraints. We show that applying different stellar yield modifications does not relieve this formation timescale problem.
    The quite unrealistically short SFT required by [Mg/Fe] and total metallicity would be prolonged if a variable stellar gwIMF were assumed. Since a systematically varying gwIMF has been suggested by various observations this could present a natural solution to this problem.
    }


\maketitle


\section{Introduction}\label{sec:intro}

The stellar $\alpha$ element to iron peak element ratio, for example [Mg/Fe], is a classic indicator of the star formation timescale (SFT) for galaxies (applied in, e.g. \citealt{2005ApJ...621..673T}, \citealt{2016MNRAS.461L.102S}, \citealt{2016ApJ...818..179L}, \citealt{2019ApJ...880L..31K}, and \citealt{2019MNRAS.490..848D}). It is known that possibly more than half of the iron in the Universe is produced by Type Ia supernovae (SNIa) which explode on average about a billion years after the formation of their parental stellar population (e.g., \citealt{2012PASA...29..447M}). On the other hand, $\alpha$ elements are mainly produced by type II supernovae (SNII) which happen within tens of millions of years when the most massive stars come to the end of their lifetime. Consequently, [$\alpha$/Fe] of the gas polluted by the supernovae decreases over time starting from the highest value, that is the [$\alpha$/Fe] yield of the most massive star. When new stars are born from the cooled and recycled polluted gas, the formation time of a star is marked by its [$\alpha$/Fe]. Thus the longer the SFT of a galaxy, the lower its average stellar [$\alpha$/Fe]. With the help of galaxy chemical-evolution simulations, \citet{2005ApJ...621..673T} were able to derive an approximated relation between average stellar [$\alpha$/Fe] and the SFT of a galaxy, $t_{\rm sf}$:
\begin{equation}\label{eq:thomas_alpha_sft}
    [\alpha/\mathrm{Fe}] \approx 1/5 - 1/6 \cdot \mathrm{log}(t_{\rm sf}).
\end{equation}
This relation indicates that the averaged stellar [$\alpha$/Fe] of a galaxy (when the galaxy is 12 Gyr old) is a function of its SFT. More comprehensive consideration shows that the resulting [$\alpha$/Fe] should also depend on the metallicity of the galaxy (see below).

Since there is compelling evidence that more massive elliptical galaxies typically have a higher stellar [$\alpha$/Fe], it has been generally accepted that massive ellipticals should have shorter SFTs \citep{2004MNRAS.347..968P,2009A&A...505.1075P,2005ApJ...621..673T,2010MNRAS.404.1775T}. This is known as downsizing of the SFT\footnote{Not to be confused with the downsizing of the galaxy age or star formation rate (SFR), which is the original meaning of downsizing \citep{1996AJ....112..839C}. See \citet{2009MNRAS.397.1776F} for a summary of the many manifestations of downsizing.}, which contradicts the cosmological merger tree in the standard LCDM universe \citep{2009MNRAS.397.1776F}. We note that throughout this text, ‘downsizing’ refers to the SFT.

It seems that downsizing can be realized quite readily when assuming elliptical galaxies form in a monolithic collapse with in situ star formation dominating the stellar mass of the ellipticals and by tuning the baryonic feedback processes due to SNII and/or active galactic nucleus \citep{2004MNRAS.347..968P,2006AIPC..847..461P}. However, this only ensures that galaxies stop star formation within the required timescale but does not ensure that galaxies can grow enough stellar mass in such a short time when hydrodynamic constraints are applied. This means that the assumed gas infall rate and/or star formation efficiency used as free parameters in the model of \citet{2004MNRAS.347..968P} might be unrealistically high. Indeed, \citet[their section 5]{1994A&A...288...57M} pointed out that the dynamical free-fall time increases for more massive gas clouds and may require a merging process with higher collision velocities to achieve the mass growth of more massive ellipticals in a shorter timescale. The galactic mass--metallicity correlation also requires that such merging processes happen early, before most stars form. Otherwise, the more massive galaxy would have the same stellar chemical composition as its lower-mass component galaxies. We note that the star formation of elliptical galaxies also needs to be early, before the gas fully exchanges angular momentum and becomes a disk. Cosmological simulations struggle in fulfilling all these requirements. Under the standard hierarchical model of galaxy formation, mergers usually occur between galaxies that have already formed a stellar population. When dry mergers (without significant amounts of gas) are an important channel for the mass growth of elliptical galaxies, that is when stellar mass grows mostly from ex situ contributions, baryonic feedback during the merger does not affect the pre-existing stellar populations. Thus, the feedback is not efficient enough to cause the required level of downsizing, and the observed [Mg/Fe]--galaxy-mass relation cannot be reproduced in \citet{2005MNRAS.363L..31N} and \citet{2015MNRAS.446.3820G} when an invariant galaxy-wide stellar initial mass function (gwIMF) is assumed. But see also \citet{2009A&A...505.1075P} mimicking the monolithic formation behaviour and \citet{2011MNRAS.413L...1C} enhancing in situ star formation triggered by fly-by encounter events.

In addition to the above difficulties, there is a more imperative problem. The observed more massive ellipticals have both a higher [Mg/Fe] value and higher metallicity \citep{2010MNRAS.404.1775T,2012MNRAS.421.1908J,2014ApJ...780...33C}
However, the galaxy models which reproduce the [$\alpha$/Fe]--galaxy-mass relation do not simultaneously reproduce the metallicity--galaxy-mass relation. This problem has been recognised since 2009 by \citet{2009A&A...505.1075P} and \citet{2009MNRAS.400.1347C} and recently confirmed by \citet{2017MNRAS.464.3812F} and \citet{2018MNRAS.474.5259D} where a variable gwIMF is suggested to be necessary once [Mg/Fe] and metallicity observations are considered together. The cause of this is stated clearly both in \citet{2017MNRAS.466L..88D}, applying a semi-analytic model, GAEA, and \citet{2017MNRAS.464.4866O}, applying cosmological
hydrodynamical simulations, that is, the SFT required to reproduce the high [$\alpha$/Fe] value being too short, such that there is not enough time for gas to recycle and enrich, resulting in a reversed metallicity--galaxy-mass relation for massive galaxies. More recent studies have not solve this problem. For example, \citet{2018MNRAS.479.5448B}, following \citet{2016MNRAS.461L.102S}, show that the hydrodynamical simulation EAGLE can reproduce the [Mg/Fe]--mass relation but not the metallicity--galaxy-mass relation, and variable IMFs are suggested; while \citet{2019ApJ...880..129P} do not reproduce the trend of the [Mg/Fe]--galaxy-mass relation (cf. \citealt{2016ApJ...829L..26L}). In summary, it is strongly suggested that an invariant gwIMF is not able to reproduce both the [$\alpha$/Fe]--galaxy-mass and metallicity--galaxy-mass relations, but a conclusive study is lacking.

With this contribution, we demonstrate that even without hydrodynamical considerations, that is without the complication of a non-zero gas recycling time, it is not possible to reproduce [Mg/Fe] and total metallicity simultaneously with realistic SFTs, thus strongly challenging the downsizing scheme. We point out that the estimation of SFT should depend not only on [$\alpha$/Fe], as is the case for Eq.~\ref{eq:thomas_alpha_sft}, but also on total metallicity of the stars. Massive stars with higher initial metallicity have a lower [Mg/Fe] yield because the stronger stellar wind of these stars mainly reduces the $\alpha$ element production \citep[section 2.1.3 there]{2012ceg..book.....M}. This is shown in \citet[their figure 4]{2019A&A...629A..93Y} who applied the stellar yield of massive stars from \citet[see their figure 7 to 9]{1998A&A...334..505P}. The \citet{2013ARA&A..51..457N} stellar yield also agrees with this trend. We note that the aforementioned studies only provide stellar yields with stellar initial metallicities [Z]=log$_{10}$(Z/Z$_\odot$)>-2, where Z$_\odot$ is the solar metallicity. The \citet{2004ApJ...608..405C} stellar yield table extends to [Z]=-4 and Z=0, suggesting a reversed trend when [Z]<-2. However, even if such behaviour is real, it only affects the very first generation of stars and has (if at all) a negligible effect on the galaxy chemical evolution. Thus, once the metallicity of the stellar population is considered, the estimation of the SFT is different and it becomes impossible to explain the observed variation of [Mg/Fe] solely by adjusting the SFT.

This paper is organised as follows. First we introduce the observations and the procedure used to estimate the SFT in Sections~\ref{sec:data} and \ref{sec:Method}, respectively. Our result, fitting only the [Mg/Fe] observation, reproduces the downsizing relation suggested by previous studies (this validates our chemical evolution code), while the result from fitting both [Mg/Fe] and total metallicity suggests a more severe downsizing relation, as is shown in Section~\ref{sec:Result}. Possible variations on the adopted stellar yield table, star formation history (SFH), and the gwIMF are discussed in Section~\ref{sec:Uncertainty}. We conclude that even with a modified stellar yield the required SFT is either too short for the massive galaxies or too long for the low-mass galaxies, compared to independent SFT constraints summarized in Section~\ref{sec:other_method}. Finally, we conclude in Section~\ref{sec:conclusion} that a systematic variation of the gwIMF appears to be a promising solution. In line with previous studies \citep{1996ApJS..106..307V,1997ApJS..111..203V,2009A&A...499..711R,2009MNRAS.400.1347C,2013MNRAS.435.2274W,2014MNRAS.442.3138F,2015MNRAS.446.4168R,2015MNRAS.449.1327V,2015MNRAS.446.3820G,2017MNRAS.464.3812F,2018MNRAS.474.5259D}. 

A comprehensive discussion of the likelihood of variable IMF formulations given the metallicity/[Mg/Fe]--galaxy-mass relation (considering possible systematic observational errors in galaxy element abundances, see Section \ref{sec:data}) is performed in our accompanying paper \citep[in prep.]{2020Yan}.

\section{Data}\label{sec:data}

\subsection{[Z/X]--galaxy-mass relation}

\begin{figure}
    \centering
    \includegraphics[width=\hsize]{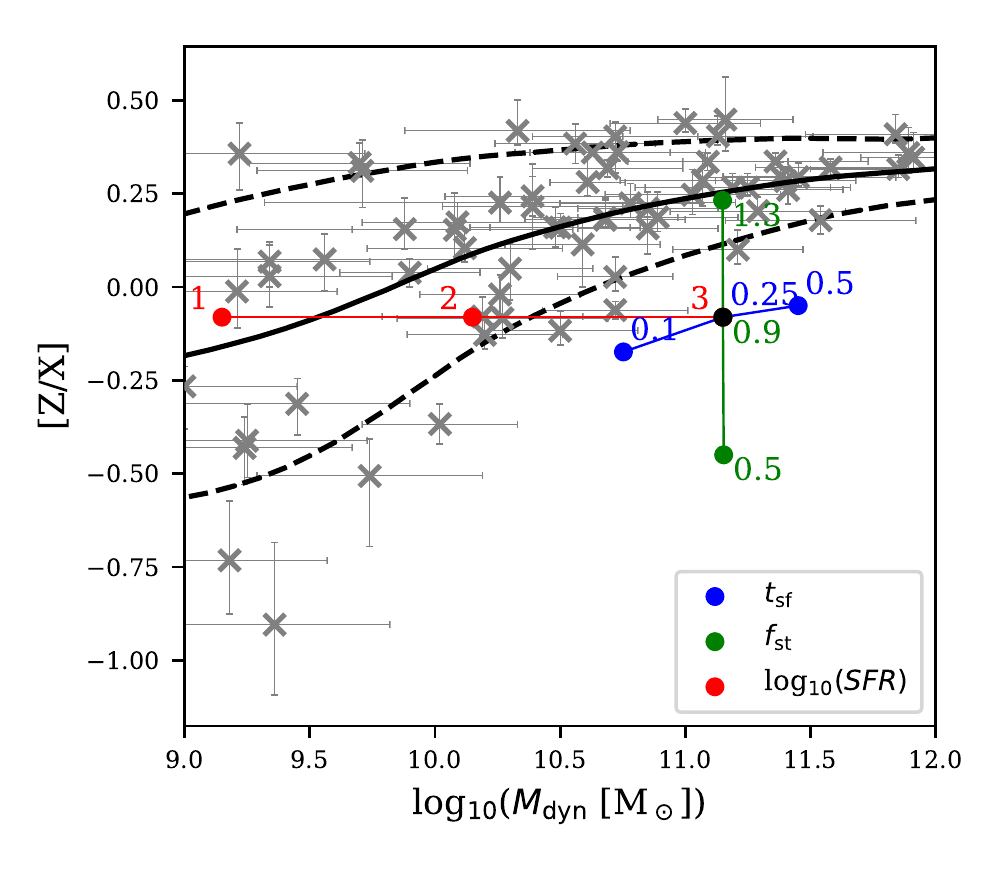}
    \caption{Observed relation between galaxy mass and metallicity, [Z/X] for local E- and S0-type galaxies. The [Z/X] data adopted from \citet{2010MNRAS.402..173A} are shown as crosses. The mean and standard deviation derived from the data are given as the solid and the dotted lines (see Section~\ref{sec:data}). Modeled stellar-mass-weighted [Z/X] of galaxies at 14 Gyr with a given SFT, $t_{\rm sf}$ (0.1, 0.25, or 0.5 Gyr), accumulated star formation fraction, $f_{\rm st}$ (0.5, 0.9, or 1.3), and SFR, $SFR$ (10, 100, or 1000 M$_\odot$/yr), assuming the invariant canonical gwIMF, are plotted with coloured filled circles. The results are shown when varying only one of the above parameters in comparison with the black filled circles.}
    \label{fig:ZX_obs}
\end{figure}

We adopt the typical metallicity--galaxy-mass relation of E- and S0-type galaxies from \citet[their table B1]{2010MNRAS.402..173A}. The notation of metallicity is changed from [Z/H] to [Z/X] following \citet[their equation 6]{2019A&A...629A..93Y}. We note that the [Z/X]--galaxy-mass relation depends on the galaxy type. A lower mean metallicity is suggested when considering all types of galaxies \citep{2005MNRAS.362...41G}. The data of single galaxies are shown by the crosses in Fig.~\ref{fig:ZX_obs}. They illustrate that the scatter of [Z/X] values for a given galaxy-mass is larger for lower-mass galaxies. To estimate the mean and standard deviation of the [Z/X]--galaxy-mass relation, we randomly generate 1000 realizations for each data point assuming that the observational error has a normal distribution on the [Z/X] value. When the given error is asymmetric, we assume the probability distribution above and below the data value is half a normal distribution. We then calculate the mean and the standard deviation of the metallicity with a given galaxy mass using all the realized galaxies that have a mass within $\pm0.5$ dex of the given galaxy mass, where the 0.5 dex is approximately the typical mass uncertainty for a single galaxy. The results are shown by the solid and dashed lines in Fig.~\ref{fig:ZX_obs}.

\subsection{[Mg/Fe]--galaxy-mass relation}

\begin{figure}
    \centering
    \includegraphics[width=\hsize]{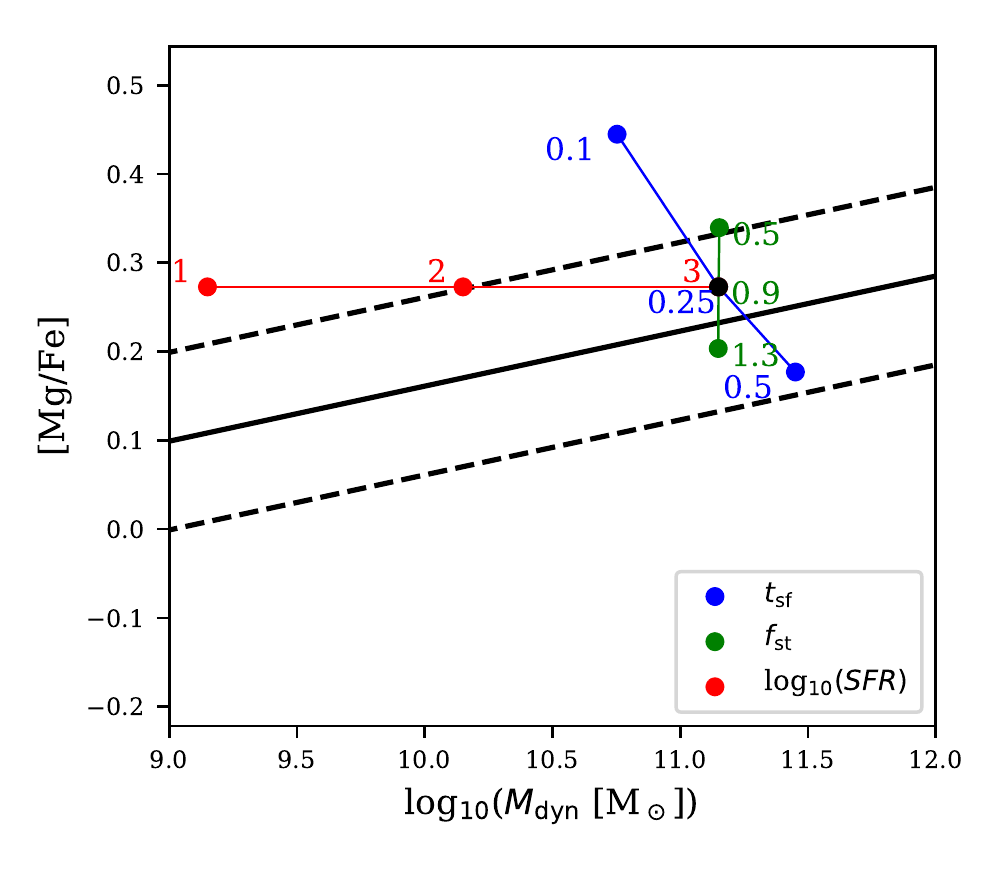}
    \caption{The [Mg/Fe]--galaxy-mass relation given in \citet{2005ApJ...621..673T} (solid line) and the assumed 0.1 dex standard deviation region (the dashed lines, see Section~\ref{sec:data}). Coloured filled circles are the same as in Fig.~\ref{fig:ZX_obs} but for [Mg/Fe] values.}
    \label{fig:MgFe_obs}
\end{figure}

The procedure of calculating the mean and standard deviation of [Mg/Fe] values of galaxies with different masses is simpler as it can be well described by a linear relation with a negligible fitting error. See for example the data in \citet{{2010MNRAS.402..173A}} and an assembly of data in \citet{2016ApJ...829L..26L} where the [Mg/Fe] values have a typical observational error of 0.05 dex for a given galaxy and a standard deviation of less than 0.1 dex for galaxies with a given mass (when the galaxy mass is larger than about $10^{9.5}$ M$_\odot$). 
In order to compare our results with the previous SFT study, we adopt the [Mg/Fe]--galaxy-mass relation of \citet[their equation 3]{2005ApJ...621..673T} as the mean and assume a standard deviation of 0.1 dex as is shown by the dashed lines in Fig.~\ref{fig:MgFe_obs}. Thus the [Mg/Fe] data of single galaxies are neither used nor plotted. We note that this assumption is only made to reproduce the work done by \citet{2005ApJ...621..673T} as detailed in Section~\ref{sec:reproduce} below. Adopting a different [Mg/Fe]--galaxy-mass relation, for example shifting the relation vertically, would not affect our main conclusions (cf. Section~\ref{sec:yield_uncertainty}).

We note that there is a systematic observational uncertainty due to the different assumptions applied in the stellar synthesis model (stellar evolution and atmosphere model). This is shown for example by the mismatch of the intercept of the fit between five different studies compared in \citet[their figure 18]{2014ApJ...780...33C}, which can be as large as 0.1 dex. This indicates that the slope, rather than the intercept, of the galaxy-mass--[Mg/Fe] relation contains the most reliable information. Fortunately, it is also the case that the intercept of this relation is uncertain in galaxy evolution models due to the larger uncertainty of stellar yields, while the slope is largely unaffected under different yield assumptions (see Section \ref{sec:stellar_yield}). Thus, it is possible to unearth some truth about the galaxies even when the observational metal abundance and the stellar yields are both unsettled.

\subsection{Dynamical mass}\label{sec:Dynamical_mass}

To compare the chemical composition of the observed galaxies and our galaxy models for a given galaxy mass, we first need to clarify the definition of the galaxy mass where the data is adopted from, that is \citealt{2010MNRAS.402..173A} and \citealt{2005ApJ...621..673T}, which we discuss separately in the following paragraphs.


\citealt{2010MNRAS.402..173A}, from whom we adopted the metallicity--galaxy-mass relation, estimated the mass of a galaxy within the effective radius from the velocity dispersion in the galactic centre and effective radius using the virial theorem (see also \citealt{1997AJ....114.1365B} and \citealt{2000AJ....120..165T}). This is defined as the dynamical mass of a galaxy, $M_{\rm dyn}$, which is independent of gwIMF as pointed out by \citet[their section 3.7]{2010MNRAS.404.1775T}. The estimation of $M_{\rm dyn}$ in \citet[their equation B1]{2010MNRAS.402..173A} using central velocity dispersion assumes the luminosity profile of the ellipticals (cf. \citealt[their equation 4.249b]{2008gady.book.....B}) in which dark matter and gas contribute a mass portion within the effective radius from (i) negligible (e.g. \citealt[their section 4.9.2]{2008gady.book.....B} and \citealt[their figure 6]{2019ApJ...877...91B}) to (ii) a few tens of percent (e.g. \cite{2011MNRAS.415..545T}). In case (i), assuming that the stellar remnant mass profile follows the stellar mass profile, and both follow the luminosity profile, the observational $M_{\rm dyn}$ is approximately the mass of living stars and stellar remnants within the effective radius. Thus, the $M_{\rm dyn}$ in \citet{2010MNRAS.402..173A} is comparable with the $M_{\rm mod}$ of our model, which is the total mass of living stars and stellar remnants of a modelled galaxy at 14 Gyr. On the other hand, in case (ii), $M_{\rm dyn}$ is larger than $M_{\rm mod}$ and they cannot be compared directly. Therefore, we also checked that defining $M_{\rm mod}$ to be ten times larger than the current definition (i.e. ten times the mass of living stars and stellar remnants), that is to shift our calculation results (filled circles in Fig.~\ref{fig:ZX_obs} and \ref{fig:MgFe_obs}) horizontally by 1 dex, would lead to insignificant changes in our results in the following sections. If we were to set $M_{\rm mod}$ to be one hundred times the mass of living stars and stellar remnants in the model, the results in the following sections would be different but the general trend would be preserved and so would the conclusions that follow. Thus, the discussion below is robust under different dark matter and gas mass portions.

It is possible to further simplify the mass estimation (with the help of certain gwIMF assumptions) by applying the empirical Kormendy relation (\citealt{1977ApJ...218..333K}; e.g. \citealt{2009A&A...499..711R}) or, equivalently, using the Faber-Jackson relation (\citealt{1976ApJ...204..668F}; e.g. \citealt[their equation 2]{2005ApJ...621..673T}) such that the galaxy mass is solely a function of central velocity dispersion. From \citet[their equation 3]{2005ApJ...621..673T} we adopt the [Mg/Fe]--galaxy-mass relation. The "stellar mass", $M_{\rm *,tot}$, in this given relation actually includes stellar remnants, as they are part of the mass-to-light ratio calculated for a stellar population, hence $M_{\rm *,tot}$ is equal to our $M_{\rm mod}$ and we can safely apply \citet[their equation 3]{2005ApJ...621..673T} and replace the $M_{\rm *,tot}$ symbol with $M_{\rm dyn}$.

\section{Method}\label{sec:Method}

\subsection{Calculation of galaxy chemical evolution}\label{sec:evolution_model}

In our fiducial model, we assume the monolithic collapse formation scenario of elliptical galaxies and a gwIMF which has the same shape as the canonical invariant IMF given in \citet[their equation 2]{2001MNRAS.322..231K}. The simulation is performed with the open-source galaxy chemical evolution model developed by \cite{2019A&A...629A..93Y} which allows pre-specification of the SFH (cf. \citealt{1999MNRAS.302..537T}). This scheme adopts the stellar yield tables from, \citet{2001A&A...370..194M} and \citet{1998A&A...334..505P} for AGB and massive stars, respectively, SNIa yields from \citet[their TNH93 dataset]{1997MNRAS.290..623G}, the delay time distribution of SNIa given by \citet{2012PASA...29..447M}, no gas flows, and instantaneous-well-mixing of the gas, that is we treat the galaxy as an isolated point.

For each galaxy model, we specify a different initial gas mass and a SFH defined by some constant SFR, $SFR$, over a given SFT, $t_{\rm sf}$, that is, a box-shaped SFH. The initial gas mass is determined by multiplying the summed initial mass of all the stars ever formed (according to the assumed SFH) and an accumulated star formation fraction parameter, $f_{\rm st}$. That is, the $f_{\rm st}$ parameter is defined as the ratio between the accumulated stellar initial mass and the initial gas mass of the simulation. All applied parameters are listed in Table~\ref{tab:grid} which has in total $6\times7\times15=630$ possible configurations. 
\begin{table*}
    \caption{Free parameters in the model}
    \label{tab:grid}
    \centering
    \begin{tabular}{ccc}
    \hline
    Parameter [Unit] & Meaning & Applied values  \\ \hline
    log$_{10}$($SFR$[M$_\odot$/yr]) & galaxy-wide star formation rate & -1, 0, 1, 2, 3, 4, (5), (6)\tablefootmark{b}\\
    $t_{\rm sf}$ [Gyr] & galactic star formation timescale & 0.05, 0.1, 0.25, 0.5, 1, 2, 4 \\
    $f_{\rm st}$ & accumulated star formation fraction\tablefootmark{a} & 0.1, 0.2, 0.3, ..., 1.5 \\
    \hline
    \end{tabular}
    \tablefoot{
        \tablefoottext{a}{The ratio between the accumulated stellar initial mass of all the stars ever formed (according to the assumed SFH) and the initial gas mass of the simulation. This value can be larger than one since the stars return their mass into the gas phase when they die. The returned gas can then be recycled to form new stars. We note that the parameter $f_{\rm st}$ is different to the parameter $f_{\rm trans}$ defined in \citet{1999MNRAS.302..537T}. The latter is not an accumulated value and cannot be larger than one.}
        \tablefoottext{b}{Models with $t_{\rm sf}<25$ Myr and $SFR>10^4$ M$_\odot$/yr are also computed in order to cover the lower right corner of likelihood map in Fig.~\ref{fig:reproduce_1}, \ref{fig:Update_2_2}, and \ref{fig:Update_2_3} below.}
    }
\end{table*}

We note that the parameter $f_{\rm st}$ is different to the parameter $f_{\rm trans}$ defined in \citet{1999MNRAS.302..537T}. The parameter $f_{\rm trans}$ is not an accumulated value and cannot be larger than one. However, the largest possible $f_{\rm st}$ value is 1.5 if all the stellar populations return one-third of their masses instantaneously after their formation due to deceased stars, which is approximately the case when assuming the canonical gwIMF and instantaneous recycling.

The outputs of each computation are the properties of the galaxy when it is 14 Gyr old. We note that the stellar-mass-weighted property of the galaxy does not vary significantly after a few billion years as we assume a box-shaped SFH and there is no star formation activity at later times \citep{2019A&A...629A..93Y}. The results include galactic dynamical mass (mass of living star plus stellar remnant mass), $M_{\rm dyn}$, mass-weighted stellar metallicity, [Z/X], and mass-weighted stellar [Mg/Fe]. We define
\begin{equation}\label{eq:Z_over_X}
    \rm [Z/X] = log_{10}(Z/X) - log_{10}(Z_\odot/X_\odot),
\end{equation}
where $\rm Z_\odot=0.01886$ and X$_\odot$=0.70683 are the solar mass fraction of metal and hydrogen adopted from \cite{1989GeCoA..53..197A}.

An example of the output is shown in Figs.~\ref{fig:ZX_obs} and \ref{fig:MgFe_obs} (filled circles), which demonstrates the variation of the resulting [Z/X] and [Mg/Fe] values due to modifications of the initial parameters. The black filled circles are resulting from a galaxy computation with input parameters being $t_{\rm sf}=0.25$ Gyr, $f_{\rm st}=0.9$, and $SFR=1000$ M$_\odot$/yr. Coloured filled circles show how the results change when varying one of the three parameters. It turns out that varying the SFR has no effect on the chemical evolution in our fiducial model; a longer SFT reduces [Mg/Fe] as expected and mildly increases [Z/X] due to an increased number of recycling epochs (i.e. a larger number of formed stellar populations); the $f_{\rm st}$ parameter influences [Z/X] significantly and also decreases [Mg/Fe] for a higher [Z/X]. This originates from the fact that massive metal-rich stars have a lower [Mg/Fe] yield, as is mentioned above in Section~\ref{sec:intro}.

\subsection{Calculating the likelihood of each parameter set}\label{sec:calculate_likelihood}

The output values of [Z/X] and [Mg/Fe] for all models with different sets of parameters are then fitted with the observations thus determining the likelihood for each parameter configuration in Table~\ref{tab:grid}. In order to make the result smoother, we linearly interpolated the model grid calculation results of [Z/X] and [Mg/Fe] for any $t_{\rm sf}$, $f_{\rm st}$, and $M_{\rm dyn}$ combination. 

The likelihood of a given parameter set, $p(M_{\rm dyn},f_{\rm st}, t_{\rm sf})$, is calculated using the interpolated results, [Z/X]$_{\rm mod}$ and [Mg/Fe]$_{\rm mod}$, as:
\begin{equation}\label{eq:likelihood}
\begin{split}
& p(M_{\rm dyn},f_{\rm st}, t_{\rm sf}) = p_{\rm Z/X} \times p_{\rm Mg/Fe},\\
& p_{\rm Z/X} = \\
& 1-\mathrm{erf}\left[\left([\mathrm{Z/X}]_{\rm obs}-[\mathrm{Z/X}]_{\rm mod}(M_{\rm dyn},f_{\rm st}, t_{\rm sf})\right)/\sigma_{\mathrm{Z/X}}/\sqrt{2}\right], \\
& p_{\rm Mg/Fe} = \\
& 1-\mathrm{erf}\left[\left([\mathrm{Mg/Fe}]_{\rm obs}-[\mathrm{Mg/Fe}]_{\rm mod}(M_{\rm dyn},f_{\rm st}, t_{\rm sf})\right)/\sigma_{\mathrm{Mg/Fe}}/\sqrt{2}\right],
\end{split}
\end{equation}
where the erf stands for the error function, $\sigma$ is the standard deviation of the observation, and the subscript obs and mod stand for the observational value and the value interpolated from our model grid, respectively.

\section{Results}\label{sec:Result}

With a given $f_{\rm st}$, we are able to condense the $p(M_{\rm dyn},f_{\rm st}, t_{\rm sf})$ information into the $t_{\rm sf}$--$M_{\rm dyn}$ plane with the likelihood $p$ of each point presented by a colour code. The $f_{\rm st}$ is either set to be a fixed value (as in Section~\ref{sec:reproduce} and \ref{sec:update} below) or is a free parameter such that only the best-fit results are shown (as in Section~\ref{sec:update2}).

\subsection{Reproducing previously achieved results}\label{sec:reproduce}

\citet{2005ApJ...621..673T} discussed the SFT based on the $\alpha$ element enhancement of early-type galaxies (Eq.\ref{eq:thomas_alpha_sft}). As mentioned above, this earlier study considers only the [Mg/Fe] values. The assumption on the mass ratio between star and gas (i.e. $f_{\rm st}$) of \citet{2005ApJ...621..673T} is unclear according to their section 6.1. To test if our code can reproduce this previous result, we can only assume that \citet{2005ApJ...621..673T} applied an invariant $f_{\rm st}$. This is reasonable because \citet{1999MNRAS.302..537T} discussed and demonstrated mainly the [Fe/H] and [Mg/Fe] evolution and not that of [Z/X]. It is known now that the iron abundance is not a good proxy for the total metallicity of an elliptical galaxy since the [Fe/H] varies very little with the galaxy mass compared to metallicity that increases significantly with galaxy mass (e.g. \citealt{2012MNRAS.421.1908J}, \citealt{2014ApJ...780...33C}, and \citealt{2019ApJ...880L..31K}).

\begin{figure}
    \centering
    \includegraphics[width=\hsize]{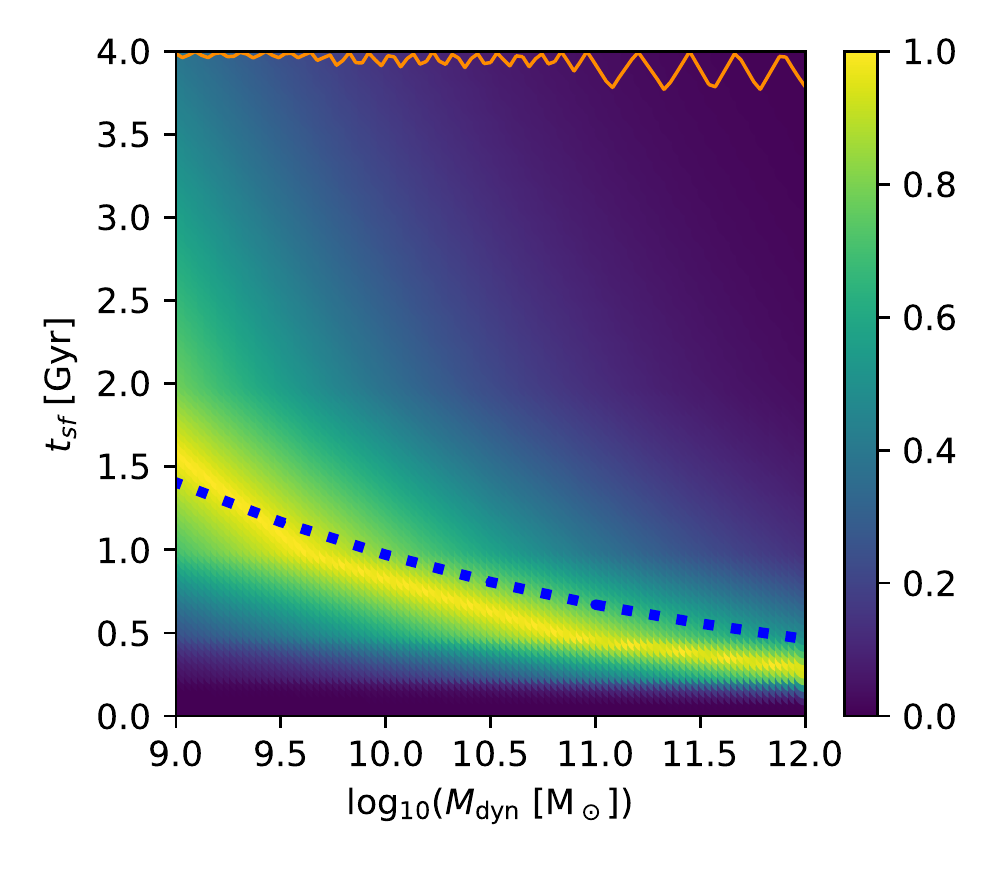}
    \caption{Likelihood (shown as the colour map) for a given SFT, $t_{\rm sf}$, and galaxy dynamical mass, $M_{\rm dyn}$, as defined in Eq.~\ref{eq:likelihood_2} assuming an invariant canonical gwIMF and fitting only the [Mg/Fe] observations. The blue dotted curve indicates the downsizing relation of \citet{2005ApJ...621..673T}. The thin orange curve is the value of the highest likelihood for any given $M_{\rm dyn}$ that follows the yellow ridge-line, sharing the y-axis with the likelihood colour-bar.}
    \label{fig:reproduce_1}
\end{figure}

To check the consistency between our model and \citet[their equation 5]{2005ApJ...621..673T}, we assume a fixed $f_{\rm st}$ value of 0.5\footnote{Fixed $f_{\rm st}$ values of 0.3, 0.4, 0.6, and 0.7 were also tested while the best fit value is 0.5.} and fit only the [Mg/Fe] observation, that is, instead of Eq.~\ref{eq:likelihood} we assume in this section:
\begin{equation}\label{eq:likelihood_2}
p(M_{\rm dyn},f_{\rm st}=0.5, t_{\rm sf}) = p_{\rm Mg/Fe}.
\end{equation}
The resulting likelihood map for galaxies with different combinations of masses and SFTs is shown in Fig.~\ref{fig:reproduce_1}. The yellow (best-fit) region reproduces the Thomas et al. relation, shown as the dashed line. Considering that there are several differences between the two models (e.g., the parameter $f_{\rm st}$ might be different, we apply a Kroupa IMF as the gwIMF while \citet{2005ApJ...621..673T} apply the Salpeter IMF as the gwIMF, the normalization and delay time distribution of SNIa are not identical, the computer codes are entirely different and independently developed, etc.), the degree of agreement is impressive.

Therefore, the galaxy chemical evolution model published in \citet{2019A&A...629A..93Y} is consistent with other existing models.

\subsection{Fitting also the metallicity}\label{sec:Consider_metallicity}

\subsubsection{Solutions with fixed star formation fraction}\label{sec:update}

\begin{figure}
    \centering
    \includegraphics[width=\hsize]{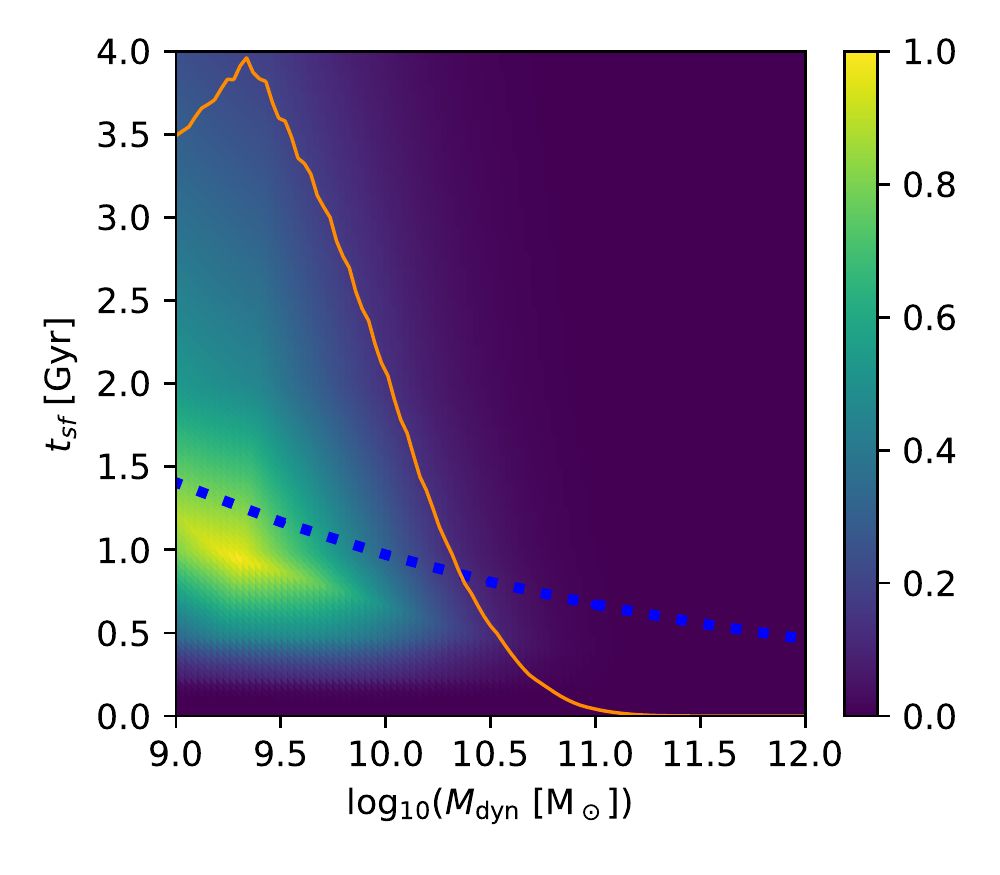}
    \caption{Same as Fig.~\ref{fig:reproduce_1} but now fitting the [Mg/Fe] and [Z/X] observations simultaneously with the likelihood calculated by Eq.~\ref{eq:likelihood} and assuming a fixed $f_{\rm st}$ value of 0.5. No solution can be found for massive galaxies since they are more metal-rich, requiring a higher $f_{\rm st}$ value (see Fig.~\ref{fig:Update_2_2} and \ref{fig:best_STF_Kroupa} below).}
    \label{fig:Update_1}
\end{figure}

Having tested that our model is in line with the previous study, we take the observational [Z/X] values into consideration. As can be expected, a model with $f_{\rm st}$ being fixed cannot reproduce the increasing [Z/X] for more massive galaxies as is shown in Fig.~\ref{fig:Update_1} (the computations for this figure all assume $f_{\rm st}=0.5$). A variable $f_{\rm st}$ is necessary.

\subsubsection{Solutions with a variable star formation fraction}\label{sec:update2}

\begin{figure}
    \centering
    \includegraphics[width=\hsize]{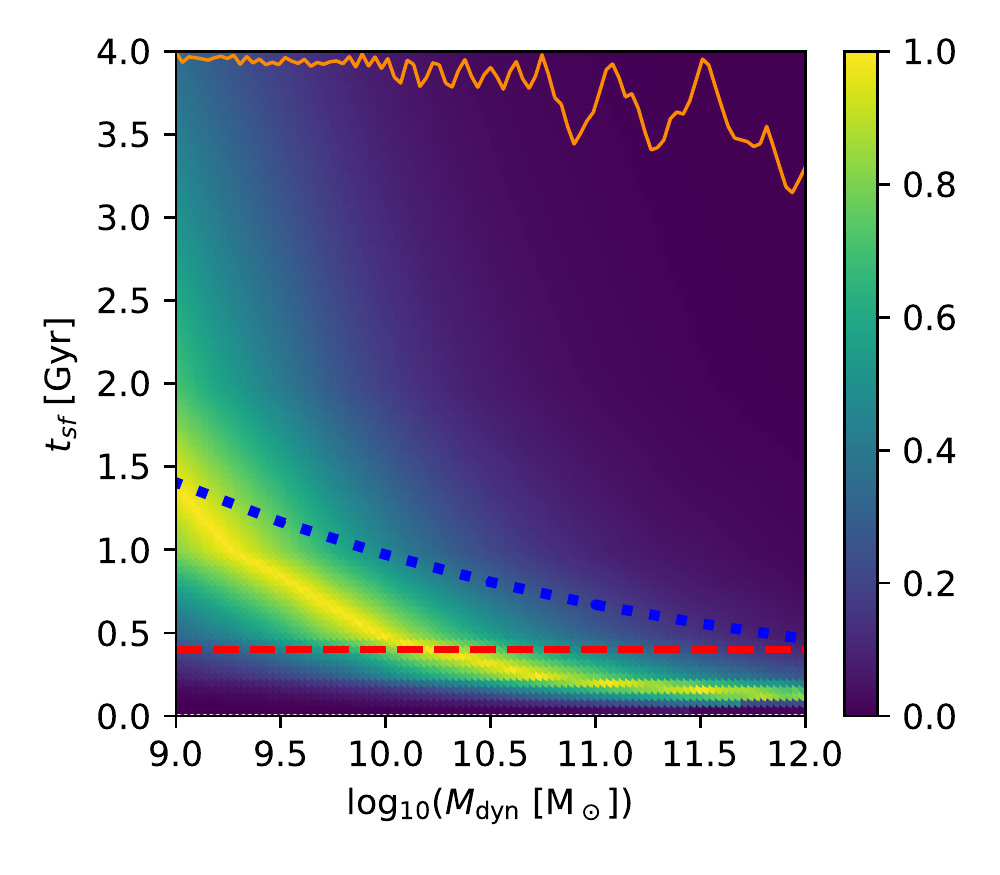}
    \caption{Same as Fig.~\ref{fig:Update_1} but with different $f_{\rm st}$ values for each galaxy mass. The $f_{\rm st}$ is shown in Fig.~\ref{fig:best_STF_Kroupa} below. It is now possible to find solutions for high- and low-mass galaxies but the required SFT for the massive galaxies are excessively short compared to the 0.4 Gyr limit given by \citet{2012MNRAS.421.1908J} shown as the red dashed line (see text).}
    \label{fig:Update_2_2}
\end{figure}

\begin{figure}
    \centering
    \includegraphics[width=\hsize]{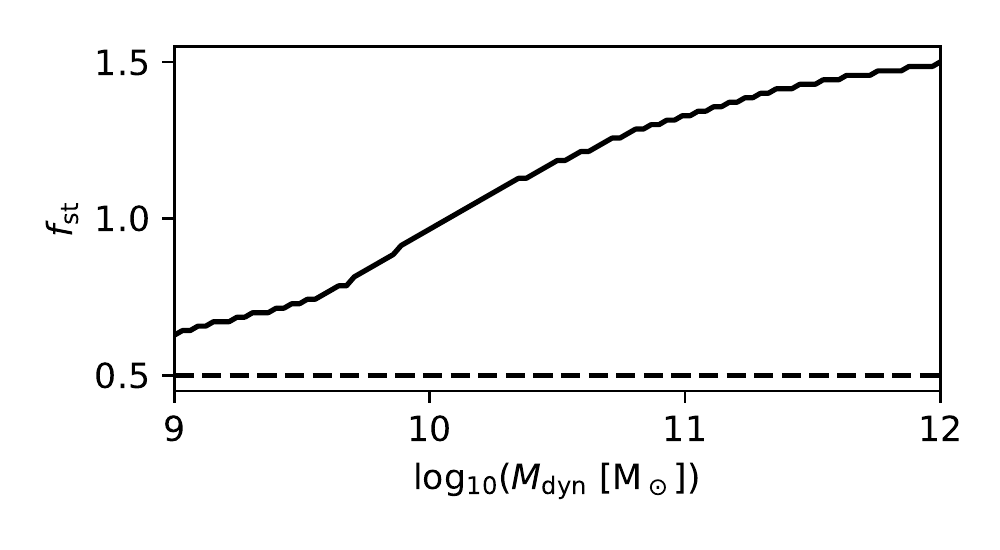}
    \caption{Constant $f_{\rm st}=0.5$ assumed for the calculations shown in Figs.~\ref{fig:reproduce_1} and \ref{fig:Update_1} (dashed line) and the corresponding accumulated star formation fraction, $f_{\rm st}$, for the best-fit solutions for each dynamical mass, $M_{\rm dyn}$, shown in Fig.~\ref{fig:Update_2_2} (solid curve). As explained in Table.~\ref{tab:grid}, $f_{\rm st}$ can be higher than one.}
    \label{fig:best_STF_Kroupa}
\end{figure}

Galaxies with a higher mass require a higher $f_{\rm st}$ value as they are more metal-rich. The most likely $f_{\rm st}$ (resulting in the highest $p$ as defined by Eq.~\ref{eq:likelihood} above) for each $M_{\rm dyn}$ is applied. The likelihood map and the corresponding $f_{\rm st}$ value for each galaxy mass are shown in Figs.~\ref{fig:Update_2_2} and Fig.~\ref{fig:best_STF_Kroupa}, respectively.

The result suggests that the SFT for massive galaxies has an even smaller value than that deduced by \citet{2005ApJ...621..673T} (shown as the dashed line). This is because massive galaxies have a higher averaged stellar metallicity while the metal-rich stars have a lower [Mg/Fe] yield. As the increasing [Z/X] ratio is now reproduced by an increasing $f_{\rm st}$, the galactic [Mg/Fe] yield in the massive galaxies becomes lower and requires an even shorter SFT than before to fit with the high [Mg/Fe] value of massive ellipticals. This is demonstrated by the green and blue filled circles in Figs.~\ref{fig:ZX_obs} and \ref{fig:MgFe_obs}.

The even shorter SFT is unrealistic in hydrodynamical simulations with a non-zero gas-recycling time (see Section~\ref{sec:intro}). It also contradicts the lower SFT limit of about 0.4 Gyr set by the observation of the [C/Mg] values as argued in \citet{2012MNRAS.421.1908J}. Thus, no solution can be found with our fiducial setup outlined in Section~\ref{sec:evolution_model}. In the following section, we drop some of the assumptions in order to find a solution.

\section{Possible modifications of the fiducial model}\label{sec:Uncertainty}

Here we explore whether the above result is robust under different assumptions on the stellar yield table and the IMF.

Since the first report of the [Mg/Fe]--galaxy-mass relation \citep{1992ApJ...398...69W}, two alternative explanations have been proposed: a selective galactic wind, or a change of the gwIMF. The selective galactic wind is difficult to study theoretically. It is also disfavoured by \citet{1994A&A...288...57M} because most stellar populations would have already formed before the onset of strong galactic wind. Nevertheless, the idea can be explored with ad-hoc assumptions (e.g. \citealt{2013MNRAS.435.3500Y}). On the other hand, a modification of the gwIMF has been expected and discussed for a long time \citep{1999MNRAS.302..537T,2005MNRAS.363L..31N,2009A&A...499..711R,2010MNRAS.408..272S,2015MNRAS.446.3820G,2016MNRAS.456L.104M,2017MNRAS.464.3812F,2018MNRAS.475.3700M,2018MNRAS.479.5448B}, which, as we argue below, is the most promising solution.

Other possible modifications of the model such as more sophisticated SFHs, alternative delay time distributions (DTDs) of SNIa events, and stellar rotation speed considerations are beyond the scope of the current study and we do not discuss these for now.

\subsection{Modifying the stellar yield table}\label{sec:stellar_yield}

The stellar total yield of element $i$ in our model is assumed to be a function of the stellar initial mass and its metallicity, $y_{i}(Z_*, M_*)$. For example, the element yield of a star with given mass and metallicity can be expressed as
\begin{equation}\label{eq:yield}
y_{ijk}(M_*=M_j, Z_*=Z_k)=a_i+b_{ij}\cdot M_j+c_{ik}\cdot Z_k + {\mathcal {O}}(M_j, Z_k),
\end{equation}
where ${\mathcal {O}}$ represents the higher-order terms.

Sections \ref{sec:yield_uncertainty} and  \ref{sec:yield_uncertainty_2} discuss the cases of modifying parameters $b_{ij}$ and $c_{ik}$, respectively.

\subsubsection{As a function of stellar initial mass, $M_*$}\label{sec:yield_uncertainty}

If the element-yields--stellar-mass relation is modified (e.g. parameter $b_{ij}$ in Eq.~\ref{eq:yield}), the stellar-population-averaged yield can be different, but the change would be the same for all galaxies since they have the same gwIMF. The modification can then be accounted for by a single adjustment of the stellar-population averaged yield, which alters all the [Mg/Fe] results (e.g. thin coloured lines in Fig.~\ref{fig:MgFe_obs}) and/or all the [Z/X] results (e.g. thin coloured lines in Fig.~\ref{fig:ZX_obs}) by the same amount. The modifications of the [Mg/Fe] and [Z/X] results could be similar when the Mg yield is modified since it is linked with the oxygen yield which is the most abundant metal element. As we verified, ignoring the [Z/X] modification does not change our conclusions because [Z/X] affects mostly the determination of the $f_{\rm st}$ parameter rather than $t_{\rm sf}$.

As mentioned in Section~\ref{sec:update2}, SFT should be at least 0.4 Gyr for all galaxies. To increase the required SFT values, a higher galactic IMF-weighted [Mg/Fe] yield is needed. The theoretical stellar Mg yields can indeed be higher than the current applied values \citep[their section 9.7]{1998A&A...334..505P} by as much as 0.3 dex \citep{1998MNRAS.296..119T,2005A&A...436..879K,2018ApJ...861...40P}. Assuming that the stellar yield difference has a linear effect on the galactic yield, the gwIMF-weighted [Mg/Fe] yield might be higher also by about 0.3 dex.

\begin{figure}
    \centering
    \includegraphics[width=\hsize]{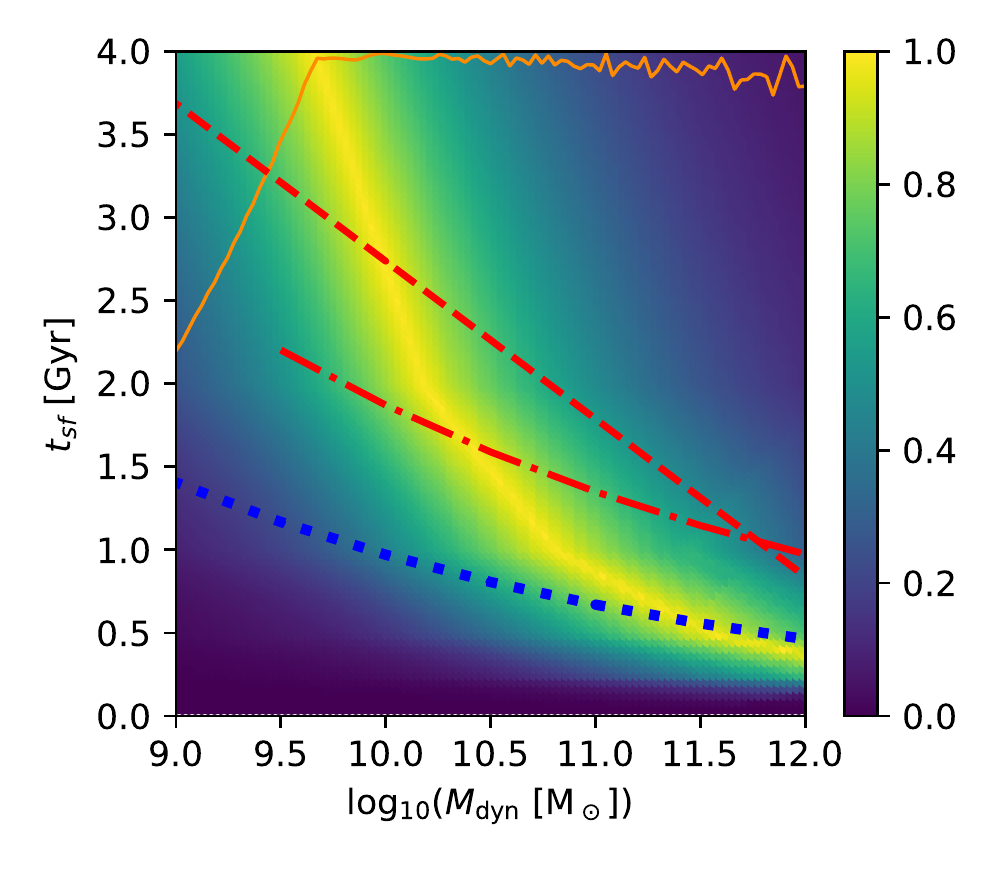}
    \caption{Same as Fig.~\ref{fig:Update_2_2} but assuming a higher galactic Mg yield by 0.15 dex (40\%) increase. The additional red dashed line and the red dash-dot line are the half-mass formation timescale determined with the stellar population synthesis method, independent of our chemical evolution constraints, given by \citet{2011MNRAS.418L..74D} and \citet{2015MNRAS.448.3484M}, respectively. We note that the red lines are rough estimations (see Section~\ref{sec:other_method} below) and may shift horizontally by 1 or 2 dex depending on the amount of gas and dark matter involved in the estimation of $M_{\rm dyn}$ (see Section~\ref{sec:Dynamical_mass}), which would not affect our conclusions.}
    \label{fig:Update_2_3}
\end{figure}

With a higher [Mg/Fe] yield, the best-fitting SFTs of massive ellipticals can reach 0.4 Gyr. However, the best-fitting SFT for the low-mass galaxies increases more dramatically. This is shown in Fig.~\ref{fig:Update_2_3} with a model assuming a galactic [Mg/Fe] yield higher by 0.15 dex than the fiducial model (while the [Z/X] galactic yield is not modified). 
The best-fitting SFTs for low-mass galaxies become much longer than other independent observational estimations (see Fig.~\ref{fig:Update_2_3} and Section~\ref{sec:other_method} below), suggesting that the required SFT  to reproduce the [Mg/Fe] observations in such a model is not fulfilled by real galaxies.

In summary, we find no solution which satisfies the SFT constraints of massive and low-mass galaxies simultaneously solely by modifying the stellar yield table as a function of the stellar initial mass. It might be possible to find a solution by combining the stellar yield modification with a specific galactic-wind-strength--galaxy-mass relation, but there will be parameters to be fine-tuned while other explanations could be simpler.

\subsubsection{As a function of stellar initial metallicity}\label{sec:yield_uncertainty_2}

If the metal-rich stars produce more magnesium than the currently applied value, it is in principle possible to increase the required SFT of massive galaxies that have a higher metallicity while not affecting the low-mass galaxies. 
It has been suggested that metallicity may correlate with the rotation speed of the stars which leads to a different yield, which seems to be in line with this scenario. The consideration of yields modified by stellar rotation speed is very uncertain, but \citet{2019arXiv190709476R} demonstrated a useful approach to this possibility.

We note that although the average stellar [Mg/Fe] and [Z/X] of galaxies in different mass bins agrees with the above ansatz, the individual low-mass galaxies, which have vastly different [Z/X] but similar [Mg/Fe], do not agree and require another explanation. This possibly involves galactic inflow and selective galactic winds. However, our current work considers the simple situation without gas flow (see \citealt{2019A&A...629A..93Y}).

In any case, we disfavour such a solution where the stellar yield is modified as a function of galaxy mass since it mimics the effect of a variable gwIMF (see the following section). These two scenarios are degenerate.

In conclusion, merely applying a different stellar yield
cannot explain the [Mg/Fe] and [Z/X] observations. The estimation of SFT when considering both observations leads to a more severe downsizing relation (such as Fig.~\ref{fig:Update_2_2} or Fig.~\ref{fig:Update_2_3}) which is not reconcilable with either hydrodynamical simulations of the formation of early-type galaxies or with independent estimations (red lines in the aforementioned figures, described in Sections \ref{sec:update2} and \ref{sec:other_method} below).



\subsection{Modifying the gwIMF}\label{sec:modify_IMF}

\subsubsection{Universal invariant non-canonical gwIMF}\label{sec:non_canonical_IMF}

If the gwIMF is universal but different from (within the observational uncertainty) the assumed one, our conclusion may be compromised.

For a steeper gwIMF which forms fewer massive stars per low-mass star, \citet{1999MNRAS.302..537T} shows that the top-light gwIMF leads to lower $\alpha$ element production and requires a shorter SFT of about $10^8$ yr to fulfil a [Mg/Fe] value of 0.2 dex for massive ellipticals. \citet{2016MNRAS.456L.104M} again, motivated by the recent evidence supporting a bottom-heavy gwIMF for massive ellipticals, shows that a steep gwIMF requires an extreme and unrealistically short SFT.

For a flatter gwIMF, \citet{2010MNRAS.402..173A} claimed that they can reproduce both [Mg/Fe] and [Z/X] observations with an invariant and slightly top-heavy gwIMF. However, their calculation adopted a wrong stellar iron yield as pointed out in the erratum declaration in \citet{2012MNRAS.424..800A}. This mistake leads to a higher [Mg/Fe] yield for stars with a higher initial metallicity, which reverses the real relation (see Section~\ref{sec:intro}) and allows (erroneously) explanation of the correlation between [Mg/Fe] and [Z/X]. With the correct stellar yield adopted, \citet{1999MNRAS.302..537T} claim that a top-heavy gwIMF has no significant effect on the simulated [Mg/Fe] value.

We note that the gwIMF of massive ellipticals is probably bottom-heavy and top-heavy at the same time \citep{2018A&A...620A..39J,2019A&A...629A..93Y} and it could have a more complicated shape. It is beyond the scope of this work to exclude a comprehensive set of possible shapes of an invariant gwIMF by simulating all the difference cases. In general however, a universal invariant gwIMF, no matter what its formulation, applies to all galaxies. Therefore, the model would encounter the same difficulty as in Section~\ref{sec:yield_uncertainty}, meaning that there is no solution for
both high- and low-mass galaxies.

\subsubsection{Variable gwIMF}\label{sec:IGIMF}

As is shown above, it is difficult to find a solution under the standard assumptions. The fact that metal-poor stars in the Milky Way have a large scatter of [Mg/Fe] values (e.g. observations compiled in \citealt{2018MNRAS.476.3432P}) and the high [Mg/Fe] value of the ultra-diffuse galaxy DGSAT I that is supposed to have an extended ($\approx 3$ Gyr) SFH \citep{2019MNRAS.484.3425M} support a non-universal gwIMF. Indeed, observational evidence is now pointing towards the IMF being dependent on the metallicity and density of molecular cloud cores on parsec scales \citep{2012MNRAS.422.2246M} and varying systematically predominantly with the SFR on galaxy scales (for recent reviews, see \citealt{2013pss5.book..115K} and \citealt{2018PASA...35...39H}).

With a top-heavy gwIMF for massive galaxies, as suggested by the IGIMF theory \citep{2013pss5.book..115K,2017A&A...607A.126Y,2018A&A...620A..39J}, it is possible to fit [Mg/Fe] and [Z/X] with a milder downsized SFT relation \citep{2009A&A...499..711R}. Not only [Mg/Fe], but also other element abundance ratios of high-redshift starburst galaxies appear to be better understood with a top-heavy gwIMF \citep{2019arXiv190806832P}. The scenario that more massive galaxies have a more top-heavy gwIMF also explains other observational properties that are difficult to understand under the invariant gwIMF scheme. For example, the early enrichment of the massive ellipticals, the systematical variation of galaxy photometric features \citep{2008ApJ...675..163H,2009ApJ...695..765M,2009ApJ...706..599L,2011MNRAS.415.1647G}, the high metal-to-star-mass ratio of massive galaxy clusters \citep{2014MNRAS.444.3581R,2017MNRAS.470.4583U}, and the systematic variation of galactic isotope abundances \citep{2017MNRAS.470..401R,2018Natur.558..260Z} readily follow if the gwIMF varies systematically as encompassed by the IGIMF theory.

Therefore, some formalism of a systematically varying gwIMF (e.g. the IGIMF theory) appears to be the most probable solution so far to explain the [Mg/Fe] and metallicity of elliptical galaxies. More importantly, when compared with other possible solutions (e.g. a fine-tuned and coordinated SFH for all galaxies; a different SNIa DTD; and stellar yield variation as a function of stellar rotation speed) the gwIMF variation provides predictions that are easier to test and to falsify. In a companion paper \citep[in prep.]{2020Yan}, we address the galaxy-mass--SFT relation given the [Mg/Fe] and [Z/X] constraints using the IGIMF theory as an alternative to the invariant canonical gwIMF applied here.

\section{Comparison with independent SFT constrains}\label{sec:other_method}

The SFT applied in galaxy evolution models must satisfy direct observational constraints even though the current constraints are neither accurate nor reliable \citep{2006ARA&A..44..141R}.

\citet{2007ApJS..171..146S,2011MNRAS.418L..74D,2015MNRAS.448.3484M} performed stellar population synthesis of local early-type galaxies based on galactic line-strength indices or full spectral fitting, supporting the downsizing of $t_{\rm sf}$ with the most massive galaxies formed on a billion-year timescale with SFTs being longer for low-mass galaxies. On the other hand, \citet{2018MNRAS.480.4379C}, who recovered the SFHs of a large ensemble of galaxies using photometric data, suggests that the SFT of local quiescent galaxies with a stellar-mass above $10^{10}$ M$_\odot$ does not correlate with stellar mass.

We note that these synthesis studies are based on an invariant gwIMF while a non-universal gwIMF would significantly change the results \citep{2013MNRAS.431..440F}. 

Also, there are some important drawbacks of this kind of method. Firstly, there is no guarantee that the SPS solution is unique. The SFH solution based on the integrated galactic light of all stellar populations always suffers from degeneracy and a prior assumption on the shape (and the smoothness) of the SFH, which may or may not be realistic, always needs to be made. Since the spectral features are dominated by the stellar population that is most luminous, the SPS method has a larger uncertainty for the modelled older stellar populations, and is therefore less reliable for galaxies with a longer SFT. Secondly, the best-fitting solution may not be realistic. Artificial simple stellar populations (SSPs) with different ages and metallicities are synthesized to fit the spectrum of the observed galaxy but there is no guarantee that these populations can be self-consistently generated under the additional constraint of needing to also satisfy the metal enrichment history of the galaxy. Finally, the abundance ratios (notably the $\alpha$-enhancement) are either not being considered, that is, the stellar population model is only a function of its age and metallicity, or when considered they have the same issue regarding the chemical evolution self-consistency. 

Therefore, the present SPS studies are affected by random and substantial systematic errors. Nevertheless, we consider these results from different groups as an envelope of solutions accounting for systematic errors. Significant SFT deviations from these results are unlikely.
The suggested SFTs from population synthesis studies are plotted in Fig.~\ref{fig:Update_2_3} as red lines. It appears that they do not agree with the suggested SFT from chemical evolution simultaneously fitting both [Mg/Fe] and [Z/X] data as mentioned above in Section~\ref{sec:yield_uncertainty} (cf. chemo-archaeological downsizing, \citealt{2009MNRAS.397.1776F}). The chemo-archaeological study yields either SFTs of massive galaxies that are too short as shown in Fig.~\ref{fig:Update_2_2} or SFTs of low-mass galaxies that are too long as shown in Fig.~\ref{fig:Update_2_3}.


Another independent approach estimating the SFT of galaxies is to count the galaxies and map the evolution with redshift of the comoving number density of quiescent elliptical galaxies \citep{2004ApJ...608..752B}. Due to the limited ability of high-redshift observations and thus selection bias as well as the possibility that a pseudo-evolution may be deduced if an incorrect cosmological model is applied \citep{2019MNRAS.488.3876B}, the constraints on SFT are weak. Besides, if the gwIMF at higher redshift were different, the estimation of the galaxy mass and SFR would be systematically biased \citep[their figure 7]{2018A&A...620A..39J}. The consensus assuming the invariant canonical gwIMF is that most massive elliptical galaxies should have formed above redshift 3, which limits the SFT to less than 2 Gyr considering the age of the Universe in consensus cosmology. Low-mass ellipticals do not have such limitations. 
To constrain the SFT instead of the ending time, one might try to count the galaxies as a function of redshift when they are still forming stars. For example, \citet{2006ApJ...644..792R} show that the distant red galaxies, which are likely to be dusty starburst precursors to elliptical galaxies, appear to be distributed evenly from redshift 1 to over 3.5. It is not trivial to interpret such observations however because the SFHs of the snapshot high-redshift galaxies are not known and they cannot be easily linked to a low-redshift counterpart. Possible galaxy mergers require additional assumptions to perform such studies. We note, however, that over the past 6 Gyr the number fraction of elliptical galaxies among all galaxies with a baryonic mass larger than $10^{10}$ M$_\odot$ remained unchanged at about 3.5\% \citep{2010A&A...509A..78D}, suggesting that mergers may not be an important concern.


In summary, the available evidence supports the idea that massive ellipticals start and finish their star formation at early times. A mildly downsizing and a no-downsizing SFT--galaxy-mass relation \citep[in prep.]{2020Yan} are both consistent with current observational constraints. On the other hand, severe downsizing, as required in the invariant gwIMF scheme, is not allowed, given the data.

\section{Conclusion}\label{sec:conclusion}

In this work, a grid of chemical evolution models is computed for elliptical galaxies with different gas-to-star transformation ratios ($f_{\rm st}$) and different SFHs (Section~\ref{sec:evolution_model}). The results are compared with observations (Section~\ref{sec:calculate_likelihood}).

First, we compare only the [Mg/Fe]--galaxy-mass relation, in which case the downsizing of SFT suggested by \citet{2005ApJ...621..673T} is reproduced (Section~\ref{sec:reproduce}). This validates our chemical evolution code \citep{2019A&A...629A..93Y}.

Furthermore, we fitted both the [Z/X] and [Mg/Fe] observations simultaneously (Section~\ref{sec:Consider_metallicity}). The results suggest a more severe downsizing relation due to the lower [Mg/Fe] yield of massive metal-rich stars. 

The uncertainty of our method and possible modifications of the applied assumptions are discussed and we conclude that, under the invariant gwIMF paradigm, a severe downsizing relation is always required no matter what stellar yield modification is applied (Section~\ref{sec:stellar_yield} and \ref{sec:non_canonical_IMF}).

Finally, we show that the severe downsizing relation is either not consistent with the SFT estimation from stellar population synthesis methods (Section~\ref{sec:other_method}) and/or it is not consistent with [C/Mg] observations and it is difficult to reconcile with hydrodynamical simulations.

Among the remaining possible solutions, we find that a systematically varying gwIMF is the most promising and well-supported (Section~\ref{sec:IGIMF}). This is detailed in the accompanying paper \citep[in prep.]{2020Yan}.

In summary:
\begin{itemize}
    \item We tested that our newly developed chemical evolution code GalIMF \citep{2019A&A...629A..93Y} is consistent with previous studies.
    \item We conclude that [Mg/Fe] as an indicator of the SFT is incorrect if [Z/X] is not taken into consideration.
    \item Furthermore, we suggest that an invariant gwIMF is unable to reproduce both [Mg/Fe]--galaxy-mass and [Z/X]--galaxy-mass observations unless an extreme downsizing relation (the yellow ridge-line in Fig.~\ref{fig:Update_2_2}) is accepted.
\end{itemize}

\begin{acknowledgements}
ZY acknowledges financial support from the China Scholarship Council (CSC, file number 201708080069). TJ acknowledges support by the Erasmus+ programme of the European Union under grant number 2017-1-CZ01- KA203-035562. The development of our chemical evolution model applied in this work benefited from the International Space Science Institute (ISSI/ISSI-BJ) in Bern and Beijing, thanks to the funding of the team “Chemical abundances in the ISM: the litmus test of stellar IMF variations in galaxies across cosmic time” (Donatella Romano and Zhi-Yu Zhang). Last but not least, we acknowledge correction and fruitful comments from Daniel Thomas.
\end{acknowledgements}

\bibliographystyle{aa} 
\bibliography{references}






\end{document}